\documentclass[runningheads]{llncs}
\usepackage{makeidx}
\usepackage[english]{babel}
\usepackage{graphicx}
\usepackage{amssymb, amsmath}
\newcommand{\ave}[1]{\langle #1 \rangle}
\newcommand{\cave}[1]{\overline{#1}}
\newcommand{\eps}{\varepsilon}
\newcommand{\keywords}[1]{\par\addvspace\baselineskip
\noindent\keywordname\enspace\ignorespaces#1}

\makeindex
\begin{document}
\frontmatter
\setcounter{page}{1}
\pagestyle{headings}  
\addtocmark{A Biologically Inspired Classifier}
\title{Biologically inspired classifier}
%
%
\author{Franco Bagnoli\inst{1,2,3} \and Francesca Di Patti\inst{2,3}}
\authorrunning{F. Bagnoli et al.}   

\index{Bagnoli, Franco}
\index{Di Patti, Francesca}
\institute{Dipartimento di Energetica, Universit{\`a} degli Studi di
Firenze,\\
via di Santa Marta 3, 50139 Firenze, Italy \and CSDC, Centro
interdipartimentale per lo Studio di Dinamiche Complesse, Universit{\`a} degli
Studi di Firenze, Italy 
\and INFN, Sez. Firenze}
\maketitle
\begin{abstract}
We present a method for measuring the distance among records 
based on the correlations of data stored in the corresponding  database
entries. The original method (F. Bagnoli, A. Berrones and F. Franci. Physica
A 332 (2004) 509-518) was formulated in the context of opinion formation. The
opinions expressed over a set of topic originate a ``knowledge network'' among
individuals, where two individuals are nearer the more similar their expressed
opinions are. Assuming that individuals' opinions are stored in a database, the
authors show that it is possible to anticipate an opinion using the
correlations in the database. This corresponds to approximating the overlap
between the tastes of two individuals with the correlations of their expressed
opinions.

In this paper we extend this model to nonlinear matching functions,
inspired by biological problems such as microarray (probe-sample pairing). We
investigate numerically the error between the
correlation and the overlap matrix for eight sequences of reference with
random probes. Results show that this method is particularly robust
for detecting similarities in the presence of traslocations.
\end{abstract}
\section{Introduction}\label{section:introduction}

Cluster analysis is used to classify a set of items into two or more
mutually exclusive  groups based on combinations of internal
variables. The goal of cluster analysis is to organize items into
groups in such a way that the degree of similarity is maximized for
the items within a group and minimized between groups.

Clustering problems arise in various domains of science, for example in opinion
formation, microarray analysis and antibody-antigens systems.

In opinion formation, one can assume that one's  opinion on a certain item is
given by the characteristics of the item, weighted by individual ``tastes''. The
tastes result from past experiences, but they do not change abruptly from time
to time. In principle, tastes can be decomposed into independent ``dimensions''.
It is rather difficult to identify such dimensions, as testified by the limited
success of market campaigns. However, it can be shown \cite{BBF2004} that
exploiting the correlations among the expressed opinions, it is possible to
deduce the distance between the tastes of two individuals. 

A DNA microarray is a collection of microscopic DNA spots of probes,  commonly
complementary to some region of a gene, arrayed on a
solid surface by covalent
attachment to a chemical matrix. DNA arrays are commonly used for
expression profiling, namely monitoring expression levels of thousands
of genes simultaneously, or for comparative genomic hybridization.
Gene expression microarray experiments can generate data sets with
multiple missing expression values. However, many algorithms for
gene expression analysis require a complete matrix of gene array
values as input, and may lose effectiveness even with a few missing
values. Methods for imputing missing data are needed, therefore, to
minimize the effect of incomplete data sets on analyses, and to
increase the range of data sets to which these algorithms can be
applied \cite{TCSBHTBA2001}. Moreover, comparison between a ``forecasted ''
value based on correlations in the dataset, and the measured one, can be
considered a consistency ``check''  of the dataset itself.

Antibodies are proteins that are used by the immune system to identify
and neutralize foreign objects, such as bacteria and viruses.
Classifying antibodies, based on the similarity of their binding to
the antigens, is essential for progress in immunology and clinical
medicine.

A striking feature of the natural immune system  is its use of
negative detection in which ``self'' is represented (approximately) by
the set of circulating lymphocytes that fail to match self. This
suggests the idea of a negative representation, in which a set of data
elements is represented by its complement set.  That is, all the
elements not in the original set are represented (a potentially huge
number), and the data itself are not explicitly stored.  This
representation has interesting information-hiding properties when
privacy is a concern and it has implications for intrusion detection. 
One of the example where this idea has been concretised is the case of
a negative database \cite{esponda06a}.

In a negative database, the negative image of a set of data records is
represented rather than the records themselves. Negative databases
have the potential  to help prevent inappropriate queries and
inferences. Under this scenario, it is desirable that the database supports
only the allowable queries while protecting the privacy of individual records,
say from inspection by an insider.  A second goal involves distributed
data, where one would like to determine privately the intersection of
sets owned by different parties.  For example, two or more entities
might wish to determine which of a set of possible ''items''
(transactions) they have in common without reveling the totality of
the contents of their database or its cardinality.

In this paper we use the microarray  example to test the introduction of
nonlinearities in the computation. Since in our model a datum is essentially
stored as the set of matching items plus the set of nonmatching ones, our
results can be applied both to positive and negative representation of data.
\section{Matching model}\label{section:matching_model} 
Let us first illustrate the problem summarizing the main results reported in
\cite{BBF2004}.

Consider a population of $M$ individuals experiencing a set of $N$ products.
Assume that each product is characterized by an $L$-dimensional array
$\boldsymbol{a}=(a^{(1)},a^{(2)},\ldots,a^{(L)})$ of features, while each
individual has the corresponding list of $L$ personal tastes on the same
features $\boldsymbol{b}=(b^{(1)},b^{(2)},\ldots,b^{(L)})$. The opinion of
individual $m$ on product $n$, denoted by $s_{m,n}$,  is defined proportional to
the scalar product between $\boldsymbol{b}_m$ and $\boldsymbol{a}_n$:
$s_{m,n}=\lambda(L)\; \boldsymbol{b}_m\cdot{\boldsymbol{a}_n}$, where
$\lambda(L)$ is a suitably chosen normalization factor. In general, $\lambda(L)$
should scale as $L^{-1}$ and depend on the ranges of $\boldsymbol{a}$ and
$\boldsymbol{b}$. 

In order to predict whether the person $j$ will like or dislike a certain
product $\boldsymbol{a}_n$, \emph{assuming to know $\boldsymbol{a}_n$}, it is
sufficient to obtain the individual tastes of that individual, i.e. the vector
$\boldsymbol{b}_j$. The similarity between tastes of two individuals $i$ and $j$
is defined by the overlap $\Omega_{ij}=\boldsymbol{b}_i\cdot\boldsymbol{b}_j$
between the preferences $\boldsymbol{b}_i$ and $\boldsymbol{b}_j$. 

One can build a knowledge network among people, using the vectors
$\boldsymbol{b}_m$ as nodes and the overlaps $\Omega_{ij}$ as edges. Maslov and
Zhang~\cite{MZ2001} (MZ) assume that a fraction $p$ of these overlaps are
known.  They show that there are two important thresholds for $p$ in order to be
able to reconstruct the missing information. The first one is a percolation
threshold, reached when the fraction of edges $p$ is greater than $p_1=1/M-1$
where $M$ is the number of people. This means that there must be at least one
path between two randomly chosen nodes, in order to be able to predict the
second node starting from the first one.

Since vectors $\boldsymbol{b}_n$ lie in an $L$ dimensional space, and a single
link ``kills'' only one degree of freedom, a reliable prediction needs more than
one path connecting two individuals. Maslov and Zhang show that there is a
``rigidity'' threshold $p_2$, of the order of $2L/M$, such that for $p>p_2$ the
mutual orientation of vectors in the network  is fixed, and the knowledge of the
preferences of just one person is sufficient to reconstruct those of all the
other individuals.

In general one does not have access to  individuals' preferences, nor one knows
the dimensionality $L$ of this space. In order to address  this problem, the
authors define the correlation  $C_{ij}$ between the opinions of agents $i$ and
$j$ by 
\begin{equation}\label{corr} C _{ij}=\frac{\sum_{n=1}^{N}
(s_{in}-\cave{s}_i)(s_{jn}-\cave{s}_j)}
{\sqrt{\sum_{n=1}^{N}(s_{in}-\cave{s}_i)^{2}\sum_{n=1}^{N}(s_{jn}
-\cave{s}_j)^{2}}}, \end{equation} 
where $\cave{s}_i$ is the average of the
opinion matrix  $S$ over column $i$. The elements $C _{ij}$ can be 
conveniently stored in a $M \times M$ opinion correlation matrix $C$.

One can compute an accurate opinion anticipation
$\tilde{s}_{mn}$  of a true value $s_{mn}$ using this formula:
\begin{equation}\label{estim}
\tilde{s}_{mn}=\frac{k}{M}\sum_{i=1}^{M} C_{mi}s_{in}
\end{equation}
where $k$ is a factor that in general depends on $L$ and 
on the statistical properties of the hidden components. 
However, if the components of
$\boldsymbol{a}_{n}$ and $\boldsymbol{b}_{m}$ are independent random variables,
$k$ is independent of $n$ and $m$, so it
can be simply chosen in order to 
have $\tilde{s}_{mn}$ defined over the same interval as $s_{mn}$. 

For large values of $N$ and $M$, the factor $k$ can be identified  with
the number of components $L$, and obtain an estimate for the average
prediction error
\begin{equation}\label{error}
	 \eps = \sqrt{\frac{1}{MN}\sum_{mn} \left(\tilde{s}_{mn} -
	{s}_{mn}\right)^2} \simeq \gamma L^{3/2}
	\frac{\sqrt{M}+\sqrt{N}}{\sqrt{MN}},
\end{equation}
where
\begin{equation}\label{gamma}
\gamma = \lambda(L) \sqrt{\ave{a^{2}}\ave{b^{2}}}.
\end{equation}
Formula~(\ref{error}) implies that
the predictive power of Eq.~(\ref{estim}) grows with $MN$ and diminishes
with $L$. This fact is a consequence of the decay of the correlations among
opinions with $L$, so that more amount of information is needed in order
to perform a prediction as $L$ grows. This condition can be compared
with the ``rigidity'' threshold $p_2$ in the MZ analysis. 
\section{Test case microarray inspired}\label{section:microarray}
In order to investigate the introduction of nonlinearities in the function
used to model the process of opinion formation, we considered the case of a
microarray.

As mentioned in section \ref{section:introduction}, microarray experiments can 
suffer from the missing values, and this fact represents a problem for many
data analysis methods, which require a complete data matrix. Although existing
missing value imputation algorithms have shown good performance to deal with
missing values, they also have their limitations. For example, some algorithms
have good performance only when strong local correlation exists in data, while
some provide the best estimate when data is dominated by global structure
\cite{GLY2006}.

Here we modified the model described in the previous section to investigate
the relationship between the correlation and the overlap between sequences.

To do this we considered an alphabet of four symbols, namely A, T, G, C,
corresponding to the four nucleotides that constitute the DNA. We used this
alphabet to generate randomly $M$ sequences of length $L$ representing the
probes of the microarray\footnote{The probes in real microarray are
discriminated generally carefully chosen in order to genes of interest.}.  Then
we generated $N$ samples of length $W$ representing the sequences to be
hybridized on the microarray.  

The correlation $C_{ij}$ between sample $i$ and sample $j$ is defined by
\begin{equation}\label{corr_microarray}
C _{ij}=\frac{\sum_{k=1}^{M} (m_{ik}-\cave{m}_i)(m_{jk}-\cave{m}_j)}
{\sqrt{\sum_{k=1}^{M}(m_{ik}-\cave{m}_i)^{2}\sum_{k=1}^{M}(m_{jk}
-\cave{m}_j)^{2}}} \qquad i,j=1, \ldots , N ,
\end{equation}
where $m_{ik}$ is the maximum complementary match between sample $i$ and probe
$k$ without gaps.

The aim is to test  the relationship between the correlation matrix $C$
and the overlap matrix $\Omega$ constructed using the following idea of 
similarity. We hypothesized to infer the similarity between sequences based on
the number of subsequences of length $L$ in common. For this reason we defined
the overlap  $\Omega_{ij}$ between sequence $i$ and sequence $j$ as the number
of subsequences of length $L$ that appear in the both sequences, divided by
$W-L+1$ for normalization. This matching function is nonlinear since the effect
of a mismatch depends on its position in the subsequence.

To test our hypothesis, we considered eight referential sequences:
\begin{description}
  \item[Seq. 0:] This is the first reference sequence,  completely random of
length $W$. 
  \item[Seq. 1:] Equal to sequence 0, except for a  mutation in the middle (this
mimics the Affimetrix central mismatch mechanism for measuring the level of
random pairing). 
  \item[Seq. 2:] Equal to sequence 0, but shifted of one basis.
  \item[Seq. 3:] Equal to sequence 0, with shift and central mutation.
  \item[Seq. 4:] First half of sequence 4 is equal to  the second half of
sequence 0, and vice versa.
  \item[Seq. 5:] First half of sequence 0 is equal to  the second half of
sequence 0,  the rest is random.
  \item[Seq. 6:] Another reference sequence.
  \item[Seq. 7:] Sequences 6 and 7 contains the same ``gene'', of length $W/3$,
in different positions. 
\end{description}
\section{Results}
To check the validity of the model described in the previous section,
we measured the error $\varepsilon_{ij}$ for the pair of sequences $i$ and $j$
defined as the absolute value of the difference between the correlation and the
overlap, namely $\varepsilon_{ij} =| C_{ij} - \Omega_{ij}|$. We performed
various simulations and than we calculated the average of the error denoted by
$\bar{\varepsilon}$.

\begin{figure}
\includegraphics{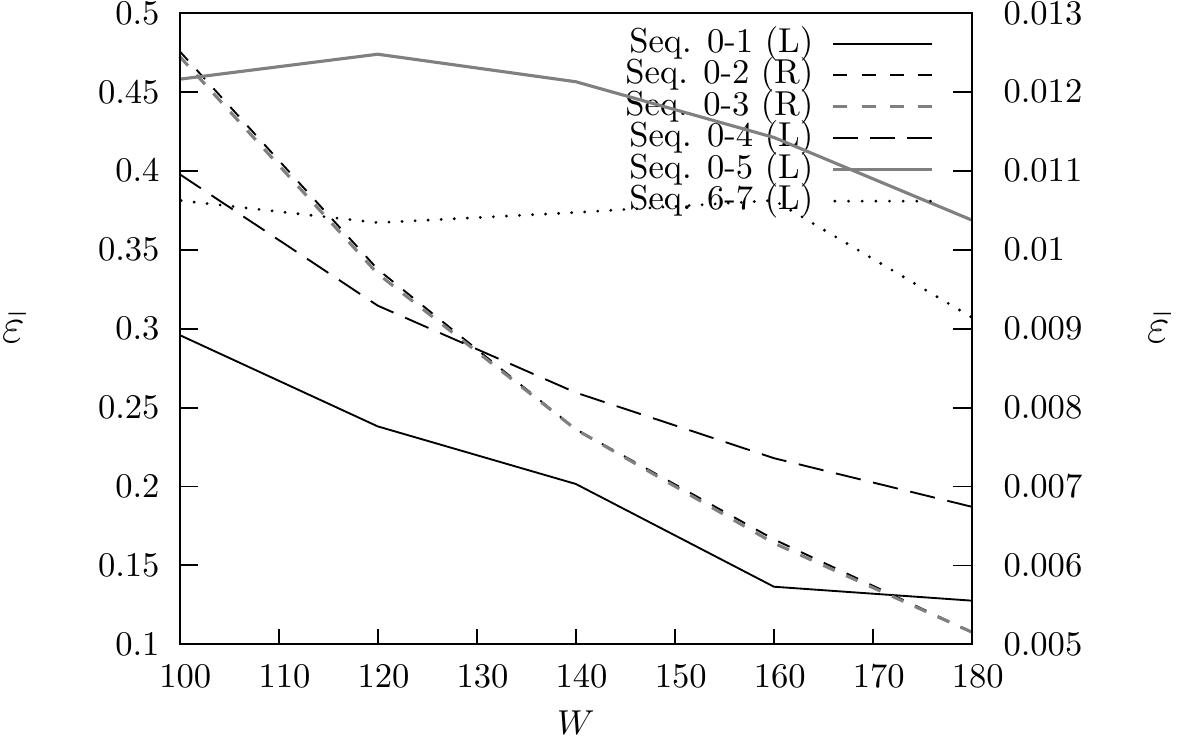}
\caption{\label{figure:mediaW}The error $\bar\varepsilon$ as a function of the
length $W$ of the samples, averaged over 40 realizations, $N=10$, $L=30$,
$M=500$. The plots of sequences 0-2 and 0-3 refer to the right y-axis. One can
observe that all errors diminish with $W$.}
\end{figure} 

In figure \ref{figure:mediaW} we plotted the error $\bar\varepsilon$ vs $W$.
One can see that for all the analysed cases the error decreases, and this
result agrees with those reported in \cite{BBF2004} (the parameter $W$ here 
corresponds to $M$ in the opinion formation model). 

\begin{figure}
\includegraphics{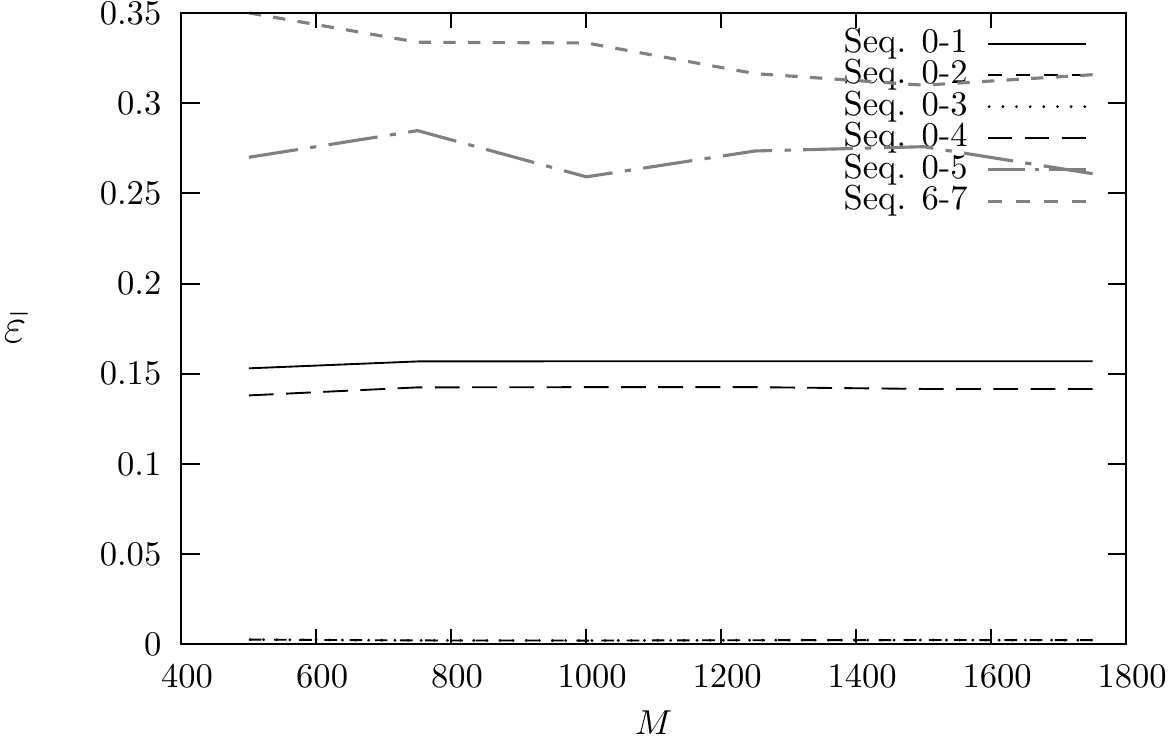}
\caption{\label{figure:mediaM}The error $\bar\varepsilon$ as a function of the
number of probes $M$, averaged over 40 realizations, $N=10$, $L=20$, $W=150$.
One can observe that errors do not vary with $M$.}
\end{figure} 

Figure \ref{figure:mediaM} shows the behaviour of $\bar\varepsilon$ with respect
to $M$. The curves are approximately constant, showing that the error
is independent of $M$.

\begin{figure}
\includegraphics{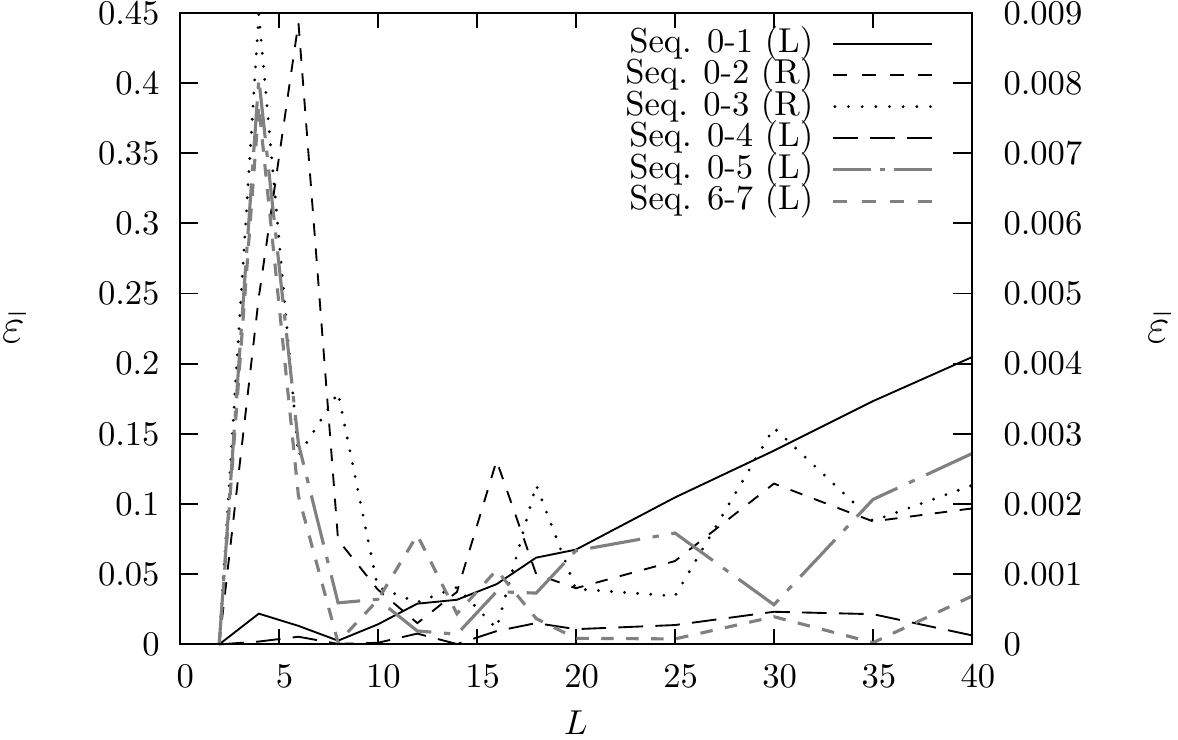}
\caption{\label{figure:mediaL}The error $\bar\varepsilon$ as a function of the
length $L$ of  the probes, averaged over 100 realizations, $N=10$, $W=200$,
$M=1000$. The plots of $\bar\varepsilon_{02}$, $\bar\varepsilon_{03}$ and
$\bar\varepsilon_{04}$ refer to the right y-axis.}
\end{figure} 

As one can see from figure \ref{figure:mediaL}, where we plotted the error
vs $L$, $\bar\varepsilon$ does not follow a monotonous trend, except
for the pair of sequences 0-4 for which the value of $\bar\varepsilon$
is almost constant and next to zero, and for $\bar\varepsilon_{01}$ which
increases. For what concerns the values of $\bar\varepsilon_{02}$,
$\bar\varepsilon_{03}$ and $\bar\varepsilon_{05}$, one can detect that the
errors decrease until $L \simeq 10$, because probes too short can hybridize in
many
positions without a high specificity. Then they oscillate until $L \simeq 35$, 
and for larger $L$ the errors increase. This last increase is due to the small
coverage of the probes in the sequence space, since we kept the number of
sequences $M$  fixed while the sequence space grows as $4^L$. 
\section{Comments}
We have proposed a method for measuring the distance among records 
based on the correlations of data stored in the corresponding  database. We
applied the method to the case of a microarray modifying the model introduced in
\cite{BBF2004} with a nonlinear matching function. More precisely, we measured
the similarity between sequences based on the number of subsequences of length
$L$ (the length of probes) in common.

We monitored the error for eight sequences of reference, with respect to $M$,
$W$, and $L$. We find that the error is low in all cases,
decreasing when $W$ increase, and independent of $M$. With
respect to $L$ we find that the model is more robust for traslocation.

In conclusion we can say that the correlation matrix of our model can be used
to
estimate the distance between sequences.  Moreover we point out that the same
result can be found following the idea of negative
database, namely using the subsequences of length $L$ not in common between two
sequences. 
\bibliographystyle{unsrt}
\bibliography{bibliography}
\end{document}